\documentclass[preprint,showpacs,pra]{revtex4}

\usepackage{graphicx}
\usepackage{dcolumn}
\usepackage{bm}
\usepackage{amsfonts}

\newcommand{\job}{J. Opt. B: Quant. Semiclass. Opt. }
\newcommand{\pr}{Phys. Rev. }

\newcommand{\jpa}{J. Phys. A }

\newcommand{\mycomm}[1]{}
\def\COMMENTS{\renewcommand{\mycomm}[1]{\par\noindent{\emph{##1}}}}

\newcommand{\UQ}{ARC Centre of Excellence for Quantum-Atom Optics, School of Physical Sciences, University of Queensland, Brisbane, Qld 4072, Australia.}

\newcommand{\etal}{{\em et al.}}
\newcommand{\e}{\mbox{e}}  

\COMMENTS

\begin{document}

\title{Bright entanglement in the intracavity nonlinear coupler}

\author{M.~K. Olsen}

\affiliation{\UQ}

\begin{abstract}

We show that the intracavity Kerr nonlinear coupler is a potential source of bright continuous variable entangled light beams which are tunable and spatially separated. This system may be realised with integrated optics and thus provides a potentially rugged and stable source of bright entangled beams.

\end{abstract}

                                                                                                                                       
\maketitle

\section{Introduction}
\label{sec:intro}

The term nonlinear coupler was given to a system of two coupled waveguides
without an optical cavity by Pe\u{r}ina \etal~\cite{coupler}.
Generically, the device consists of two parallel optical waveguides
which are coupled by an evanescent overlap of the guided modes. These may operate with both modes propagating in the same direction, or in opposite directions. The
quantum statistical properties of this device when the nonlinearity
is of the $\chi^{(3)}$ type have been theoretically investigated~\cite{Korolkova,Fiurasek},
among the predictions being energy transfer between the waveguides~\cite{Ibrahim}
and the generation of correlated squeezing~\cite{korea}. The production of entangled states of the electromagnetic field with an intracavity Kerr nonlinear coupler in which one of the modes is externally pumped was analysed by Leo\'nski and Miranowicz~\cite{Leonski}. In the travelling wave configuration, the generation of entanglement via coupled parametric downconversion has been analysed by Herec \etal~\cite{herec}. Olsen and Drummond~\cite{k2dimer} and Olivier and Olsen~\cite{nicolas} have analysed coupled intracavity downconversion in terms of continuous variable entanglement in both the above and below threshold regimes.
The system we consider here consists of two coupled nonlinear materials with $\chi^{(3)}$ nonlinearity,   
operating inside a pumped Fabry-Perot cavity. The details of the coupling are as in the work of Bache \etal~\cite{dimer}, who analysed a scheme using coupled second harmonic generation in terms of intensity correlations between the modes. The coupling is realised by evanescent overlaps
of the intracavity modes inside the nonlinear medium, which can be either a 
single Kerr crystal pumped by two spatially separated lasers, or two Kerr waveguides. This type of coupling has already been investigated both theoretically
and experimentally\cite{couple,couple2,couple3}.

In this paper we will 
calculate the phase-dependent correlations between the two
outputs of the cavity which demonstrate that the system is a source of continuous variable entangled states
and states which exhibit the Einstein-Podolsky-Rosen (EPR)~\cite{EPR} paradox. 
The spatial separation of
the entangled output modes means that they do not have to be separated by optical
devices before measurements can be made, along with the unavoidable
losses which would result from this procedure. The correlations are tunable by controlling some of the
operational degrees of freedom of the system, including the evanescent
couplings between the two modes, the input power and the cavity
detunings. The entanglement is also between outputs which can be of macroscopic intensities, which may be an advantage compared to the nondegenerate optical parametric oscillator (NDOPO) operating above threshold. Villar \etal~\cite{paulistas} recently demonstrated bright entangled outputs with such a device, but were not able to make an unambiguous demonstration of the EPR paradox due to losses. The scheme we analyse here may provide a more stable route to a demonstration with spatially separated bright beams.  

\section{Hamiltonian and equations of motion}
\label{sec:Ham}

We consider two evanescently coupled $\chi^{(3)}$ materials inside a pumped Fabry-Perot, which could be experimentally realised with, for example, a dual core fiber with dielectric mirrors at each end~\cite{Stu}. The interaction Hamiltonian in the rotating-wave approximation can be written as
\begin{equation}
{\cal H}_{int} = {\cal H}_{pump}+{\cal H}_{damp}+{\cal H}_{kerr}+{\cal H}_{couple},
\label{eq:Ham}
\end{equation}
where the pumping Hamiltonian is
\begin{equation}
{\cal H}_{pump} = i\hbar\sum_{j=1}^{2}\left(\epsilon_{j}\hat{a}_{j}^{\dag}-\epsilon_{j}^{\ast}\hat{a}_{j}\right),
\label{eq:Hpump}
\end{equation}
the damping Hamiltonian is
\begin{equation}
{\cal H}_{damp} = \hbar\sum_{j=1}^{2}\left(\Gamma_{j}\hat{a}_{j}^{\dag}+\Gamma_{j}^{\dag}\hat{a}_{j}\right),
\label{eq:Hdamp}
\end{equation}
the Kerr Hamiltonian is
\begin{equation}
{\cal H}_{kerr} = \hbar\sum_{j=1}^{2}\chi_{j}\hat{a}_{j}^{\dag\;2}\hat{a}_{j}^{2},
\label{eq:Hkerr}
\end{equation}
and the coupling Hamiltonian is
\begin{equation}
{\cal H}_{couple} = \hbar J\left(\hat{a}_{1}\hat{a}_{2}^{\dag}+\hat{a}_{1}^{\dag}\hat{a}_{2}\right).
\label{eq:Hcouple}
\end{equation}
In the above, the $\hat{a}_{j}$ are the annihilation operators for the intracavity modes, the $\epsilon_{j}$ represent the classical pumping terms, the $\Gamma_{j}$ are bath operators, the $\chi_{j}$ represent the Kerr nonlinearities, and $J$ represents the evanescent coupling. The form of the latter is as described in Bache \etal~\cite{dimer}.

As we do not expect to be able to solve the Heisenberg equations of motion, we will derive stochastic differential equations in the positive-P representation~\cite{Pplus}, after making the usual zero-temperature Markov and Born approximations for the cavity damping~\cite{DFW}. Proceeding via the usual methods~\cite{GardinerQN}, and making the correspondences between the operators $\hat{a}_{j},\:\:\hat{a}_{j}^{\dag}$ and the c-numbers $\alpha_{j},\:\:\alpha_{j}^{+}$, we find the following It\^o equations,
\begin{eqnarray}
\frac{d\alpha_{1}}{dt} &=& \epsilon_{1}-(\gamma_{1}+i\Delta_{1})\alpha_{1}-2i\chi_{1}\alpha_{1}^{+}\alpha_{1}^{2}+iJ\alpha_{2}+\sqrt{-2i\chi_{1}\alpha_{1}^{2}}\;\eta_{1},
\nonumber\\
\frac{d\alpha_{1}^{+}}{dt} &=& \epsilon_{1}^{\ast}-(\gamma_{1}-i\Delta_{1})\alpha_{1}^{+}+2i\chi_{1}\alpha_{1}^{+\;2}\alpha_{1}-iJ\alpha_{2}^{+}+
\sqrt{2i\chi_{1}\alpha_{1}^{+\;2}}\;\eta_{2},\nonumber\\
\frac{d\alpha_{2}}{dt} &=& \epsilon_{2}-(\gamma_{2}+i\Delta_{2})\alpha_{2}-2i\chi_{2}\alpha_{2}^{+}\alpha_{2}^{2}+iJ\alpha_{1}+\sqrt{-2i\chi_{2}\alpha_{2}^{2}}\;\eta_{3},
\nonumber\\
\frac{d\alpha_{2}^{+}}{dt} &=& \epsilon_{2}^{\ast}-(\gamma_{2}-i\Delta_{2})\alpha_{2}^{+}+2i\chi_{2}\alpha_{2}^{+\;2}\alpha_{2}-iJ\alpha_{1}^{+}+
\sqrt{2i\chi_{2}\alpha_{2}^{+\;2}}\;\eta_{4},
\label{eq:SDE}
\end{eqnarray}
where the $\eta_{j}$ are real Gaussian noise terms with the correlations $\overline{\eta_{j}(t)}=0$ and $\overline{\eta_{j}(t)\eta_{k}(t')}=\delta_{jk}\delta(t-t')$. We now have a set of equations which are a coupled version of those used by Drummond and Walls to describe a model of optical bistability with an intracavity Kerr medium~\cite{PeterandDan}. In principle these equations may be solved to calculate any quantity which can be written as a time-normally ordered operator product.
However, rather than immediately solving the full equations, we will linearise them about their classical steady-state solutions and examine the stability of the system. This is equivalent to treating the system as an Ornstein-Uhlenbeck process~\cite{ornstein} and allows for simple determination of the output spectra in parameter regions where the process is valid.

\section{Linearisation and stability}
\label{sec:linear}

\begin{figure}[tbhp]
\includegraphics[width=0.8\columnwidth]{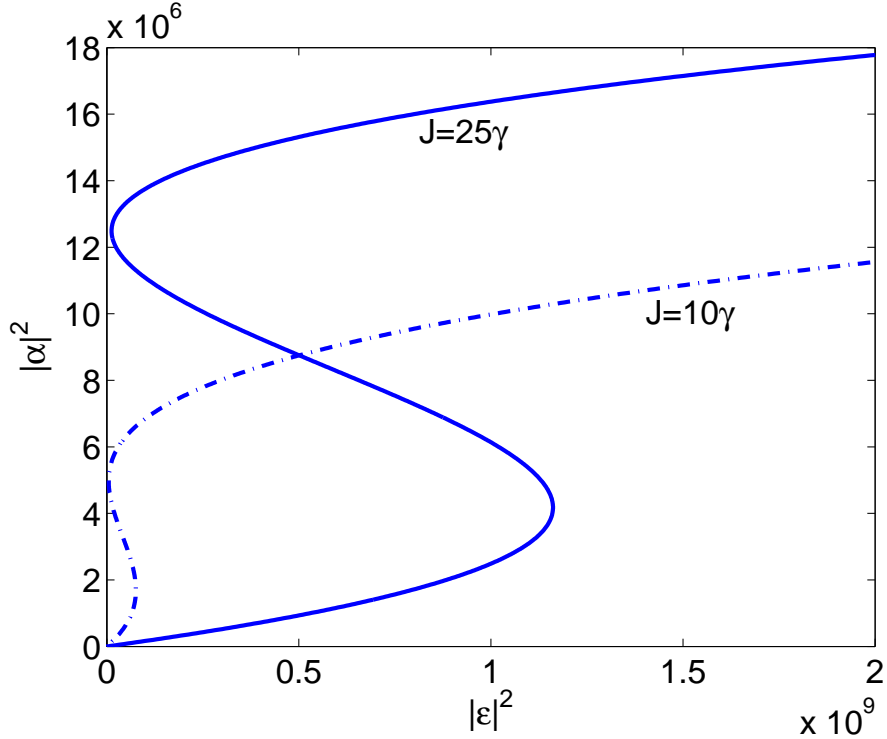}
\caption{The intensity as a function of pump intensity, for $\chi=10^{-6}$, $\gamma=1$ and $\Delta=0$, and two different values of the coupling. All quantities shown in this and subsequent figures are dimensionless.}
\label{fig:bistab}
\end{figure}

The classical equations are found by dropping the noise terms from Eq.~\ref{eq:SDE}, and are
\begin{eqnarray}
\frac{d\alpha_{1}}{dt} &=& \epsilon_{1}-(\gamma_{1}+i\Delta_{1})\alpha_{1}-2i\chi_{1}|\alpha_{1}|^{2}\alpha_{1}+iJ\alpha_{2},\nonumber\\
\frac{d\alpha_{2}}{dt} &=& \epsilon_{2}-(\gamma_{2}+i\Delta_{2})\alpha_{2}-2i\chi_{2}|\alpha_{2}|^{2}\alpha_{2}+iJ\alpha_{1}.
\label{eq:semiclass}
\end{eqnarray}
To begin with, we will consider that all the parameters are symmetric, that is $\epsilon_{1}=\epsilon_{2}=\epsilon$, $\gamma_{1}=\gamma_{2}=\gamma$, $\Delta_{1}=\Delta_{2}=\Delta$ and $\chi_{1}=\chi_{2}=\chi$. We now write equations of motion for $\alpha_{p}\;(=\alpha_{1}+\alpha_{2})$ and $\alpha_{m}\;(=\alpha_{1}-\alpha_{2})$ as
\begin{eqnarray}
\frac{d\alpha_{p}}{dt} &=& 2\epsilon-\left[\gamma+i(\Delta-J+2\chi|\alpha|^{2})\right]\alpha_{p},\nonumber\\
\frac{d\alpha_{m}}{dt} &=& -\left[\gamma+i(\Delta+J+2\chi|\alpha|^{2})\right]\alpha_{m},
\label{eq:plusandminus}
\end{eqnarray}
where we have used the symmetry of the system to set $|\alpha_{1}|^{2}=|\alpha_{2}|^{2}=|\alpha|^{2}$. The steady-state solutions of the above equations give us $\alpha_{m}^{ss}=0$, so that $\alpha_{1}^{ss}=\alpha_{2}^{ss}=\alpha=\alpha_{p}^{ss}/2$ and 
\begin{equation}
\alpha = \frac{\epsilon}{\left[\gamma+i(\Delta-J+2\chi|\alpha|^{2})\right]}.
\label{eq:alphass}
\end{equation}
Multiplying this equation by its complex conjugate and setting $\epsilon\in\Re$, we find a cubic equation for $I=|\alpha|^{2}$,
\begin{equation}
4\chi^{2}I^{3}+4(\Delta-J)\chi I^{2}+\left[\gamma^{2}+(\Delta-J)^{2}\right]I-\epsilon^{2}=0.
\label{eq:cubic}
\end{equation}
A condition for bistability is that the quadratic equation found by differentiating Eq.~\ref{eq:cubic} with respect to $I$ has two positive real roots. The quadratic is
\begin{equation}
12\chi^{2}I^{2}+8(\Delta-J)\chi I+\gamma^{2}+(\Delta-J)^{2} = 0,
\label{eq:quadratic}
\end{equation}
with roots
\begin{equation}
r_{\pm} = \frac{1}{6\chi}\left[2(J-\Delta)\pm\sqrt{(J-\Delta)^{2}-3\gamma^{2}}\right].
\label{eq:roots}
\end{equation}
The condition then is that $(J-\Delta)^{2}>3\gamma^{2}$, so that bistability is possible provided that 
\begin{equation}
J>\Delta+\sqrt{3}\,\gamma.
\label{eq:bistable}
\end{equation}
We note here that this condition is necessary but not sufficient, as it does not tell us at which values of the pumping the bistability may be found, for example. We also note that for the case where $\Delta=J$, similar to that which was used to simplify the analysis for coupled downconverters~\cite{k2dimer,nicolas}, there is no bistability. It is obvious that for the cavity on resonance (i.e $\Delta=0$), bistability is possible for any value of the coupling greater than $\sqrt{3}\gamma$, as shown in Fig.~\ref{fig:bistab}, and that for $\Delta<-\sqrt{3}\gamma$, bistability is possible for any value of the coupling. 
We will not investigate this bistability further at this time except to remark that a fully quantum analysis of an uncoupled oscillator showed that quantum fluctuations removed the bistability which was seen in a semi-classical analysis~\cite{PeterandDan}.

We will now return to the steady-state solutions themselves. Eq.~\ref{eq:cubic} may be solved analytically to give solutions which are rather complicated, so that we will proceed numerically in the general case.
However, for $\Delta=J$, the real solution simplifies to
\begin{equation}
I = \frac{-3^{2/3}\gamma^{2}+3^{1/3}\left(9\chi\epsilon^{2}+\sqrt{3(\gamma^{6}+27\epsilon^{4}\chi^{2})}\right)^{2/3}}{6\chi\left(9\chi\epsilon^{2}+\sqrt{3(\gamma^{6}+27\epsilon^{4}\chi^{2})}\right)^{1/3}},
\label{eq:iguais}
\end{equation}
which may then be substituted into Eq.~\ref{eq:alphass} to give an expression for the mode amplitudes. The values thus found agree with the steady-state mode amplitudes found by numerical integration of the equations of motion.

Now that we have found steady-state solutions, we may linearise the equations of motion around these by setting $\alpha_{j}=\alpha+\delta\alpha_{j}$ and expanding to first order in the $\delta\alpha_{j}$. This gives the equations
\begin{eqnarray}
\frac{d}{dt}\delta\alpha_{1} &=& -(\gamma+i\Delta)\delta\alpha_{1}-2i\chi\left(\alpha^{2}\delta\alpha_{1}^{+}+2|\alpha|^{2}\delta\alpha_{1}\right)+iJ\delta\alpha_{2}+\sqrt{-2i\chi\alpha^{2}}\;\eta_{1}
\nonumber\\
\frac{d}{dt}\delta\alpha_{1}^{+} &=& -(\gamma-i\Delta)\delta\alpha_{1}^{+}+2i\chi\left(\alpha^{\ast\,2}\delta\alpha_{1}+2|\alpha|^{2}\delta\alpha_{1}^{+}\right)-iJ\delta\alpha_{2}^{+}+\sqrt{2i\chi\alpha^{\ast\,2}}\;\eta_{2},\nonumber\\
\frac{d}{dt}\delta\alpha_{2} &=& -(\gamma+i\Delta)\delta\alpha_{2}-2i\chi\left(\alpha^{2}\delta\alpha_{2}^{+}+2|\alpha|^{2}\delta\alpha_{2}\right)+iJ\delta\alpha_{1}+\sqrt{-2i\chi\alpha^{2}}\;\eta_{3}
\nonumber\\
\frac{d}{dt}\delta\alpha_{2}^{+} &=& -(\gamma-i\Delta)\delta\alpha_{2}^{+}+2i\chi\left(\alpha^{\ast\,2}\delta\alpha_{2}+2|\alpha|^{2}\delta\alpha_{2}^{+}\right)-iJ\delta\alpha_{1}^{+}+\sqrt{2i\chi\alpha^{\ast\,2}}\;\eta_{4},
\label{eq:fluctuations}
\end{eqnarray}
which may be written in matrix form as
\begin{eqnarray}
\frac{d}{dt}\vec{\delta\alpha}=-A\vec{\delta\alpha}+B,
\label{eq:matriz}
\end{eqnarray}
where $\vec{\delta\alpha}=[\delta\alpha_{1},\delta\alpha_{1}^{+},\delta\alpha_{2},\delta\alpha_{2}^{+}]^{T}$, the drift matrix is
\begin{equation}
A=\left[\begin{array}{cccc}
\gamma+i(\Delta+4\chi|\alpha|^{2}) & 2i\chi\alpha^{2} & -iJ & 0 \\
-2i\chi\alpha^{\ast\,2} & \gamma-i(\Delta+4\chi|\alpha|^{2}) & 0 & iJ \\
-iJ & 0 & \gamma+i(\Delta+4\chi|\alpha|^{2}) & 2i\chi\alpha^{2} \\
0 & iJ & -2i\chi\alpha^{\ast\,2} & \gamma-i(\Delta+4\chi|\alpha|^{2})
\end{array}\right],
\end{equation}
and $BB^{T}$ is a diagonal matrix with the vector $[-2i\chi\alpha^{2},2i\chi\alpha^{\ast\,2},-2i\chi\alpha^{2},2i\chi\alpha^{\ast\,2}]$ on the diagonal. The validity of this process depends on the eigenvalues of the matrix $A$ having no negative real part, which we will verify numerically in all reported results. These eigenvalues may be found analytically as
\begin{eqnarray}
\lambda_{1,2} &=& \gamma\pm\sqrt{4\chi|\alpha|^{2}\left[3\chi|\alpha|^{2}+2(\Delta-J)\right]-(J+\Delta)^{2}},\nonumber\\
\lambda_{3,4} &=& \gamma\pm\sqrt{4\chi|\alpha|^{2}\left[2(J-\Delta)-3\chi|\alpha|^{2}\right]-(J-\Delta)^{2}}.
\label{eq:autovalores}
\end{eqnarray}
For $\Delta=J$, which we will use in the results presented below, the linearisation process is not valid for $|\alpha|^{4}>(\gamma^{2}+4J^{2})/12\chi^{2}$. 
In this region, we would need to use stochastic integration of the full equations (\ref{eq:SDE}) to obtain results.

Writing the equation in the above form (\ref{eq:matriz}) leads to a particularly simple expression for the intracavity spectral correlation matrix,
\begin{equation}
S(\omega)=\left(A+i\omega\openone\right)^{-1}BB^{T}\left(A^{T}-i\omega\openone\right)^{-1},
\label{eq:inspek}
\end{equation}
after which we use the standard input-output relations~\cite{mjc}
to relate these to quantities which may be measured outside the cavity. The matrix $S(\omega)$ contains all the information necessary to calculate the correlation functions which we will give in section~\ref{sec:measurement}.

\section{Entanglement criteria}
\label{sec:measurement}

Entanglement is a property of quantum mechanics which is related to the inseparability of the combined density matrix of a system into density matrices for its subsystems. In the present situation, we are interested in continuous variable bipartite entanglement between the modes $1$ and $2$. The presence of this entanglement can be shown in a number of ways. Before we proceed with the definition of the criteria we will use here, we will define the optical quadratures which will be used, as their actual definition affects the form of the inequalities. We define the quadrature operators at the phase angle $\theta$ as 
\begin{equation}
\hat{X}_{j}^{\theta}=\hat{a}_{j}\e^{-i\theta}+\hat{a}_{j}^{\dag}\e^{i\theta},
\label{eq:quaddef}
\end{equation}
so that $[\hat{X}_{j}^{\theta},\hat{X}_{k}^{\theta+\pi/2}]=2i\delta_{jk}$ and therefore the Heisenberg uncertainty principle (HUP) requires $V(\hat{X}_{j}^{\theta})V(\hat{X}_{k}^{\theta+\pi/2})\geq 1\delta_{jk}$. In the interests of notational simplicity, we will label the quadrature $\hat{X}_{k}^{\theta+\pi/2}$ as $\hat{Y}_{k}^{\theta}$. We also note here that our results will be presented in terms of output spectral correlations, as these are the quantities that are experimentally measurable, and that the same inequalities as presented below will hold for these. We will also present the correlation functions in terms of intracavity variances, again for reasons of notational simplicity.

\begin{figure}[tbhp]
\includegraphics[width=0.8\columnwidth]{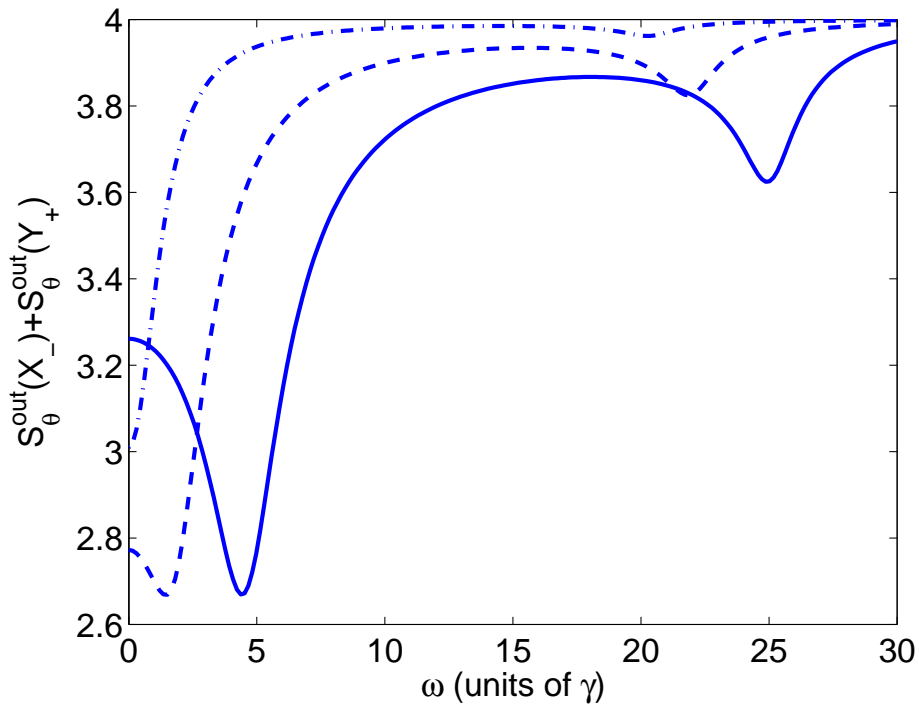}
\caption{Violation of the Duan inequalities for $\epsilon=10^{3},\gamma=1,J=\Delta=10$. The solid line is for $\chi=10^{-5}$, at a quadrature angle of $\theta=80^{\rm o}$, the dashed line is for $\chi=10^{-6}$, at an angle of $\theta=122^{\rm o}$ and the dash-dotted line is for $\chi=10^{-7}$, at an angle of $\theta=14^{\rm o}$. The inequality is violated and entanglement is demonstrated for any value less than $4$.}
\label{fig:Duan}
\end{figure}

The first of the entanglement measures is due to Duan \etal~\cite{Duan} and also Simon~\cite{Simon}, who developed inseparability criteria which are necessary and sufficient for Gaussian states, and sufficient in general. These criteria have recently been shown to be special cases of an infinite series of inequalities based on the non-negativity of determinants of matrices constructed from certain combinations of operator moments~\cite{Vogel}. In the general case, we may define the quadrature operators similarly to Duan as
\begin{eqnarray}
\hat{X}_{\pm}^{\theta} &=& |b|\hat{X}_{1}^{\theta}\pm\frac{1}{|b|}\hat{X}_{2}^{\theta},\nonumber\\
\hat{Y}_{\pm}^{\theta} &=& |b|\hat{Y}_{1}^{\theta}\pm\frac{1}{|b|}\hat{Y}_{2}^{\theta},
\label{eq:XYplusminus}
\end{eqnarray}
where $b$ is an arbitrary non-zero real number. It may be shown that, for separable states,
\begin{equation}
V(\hat{X}_{\pm}^{\theta})+V(\hat{Y}_{\mp}^{\theta})\geq 2\left(b^{2}+\frac{1}{b^{2}}\right),
\label{eq:critDuan}
\end{equation}
with any violation of this inequality therefore demonstrating the presence of bipartite entanglement. In what follows, we will choose $b=1$ so that the lower bound of the inequality is $4$. While this is not the optimal choice for the general case, it is sufficient for purposes of comparison with the other measures we will use to categorise our system. In Fig.~\ref{fig:Duan} we give the predictions for the values of the spectral output correlation $S^{out}(\hat{X}_{-}^{\theta})+S^{out}(\hat{Y}_{+}^{\theta})$ for different values of the $\chi^{(3)}$ interaction. We see that, as the interaction strength increases, the maximum violation moves away from zero frequency and is found at different quadrature angles, with the spectra becoming bifurcated. This is different from the cases with coupled $\chi^{(2)}$ interactions, where setting the detunings equal in strength to the evanescent couplings meant that the optimal correlations were found for zero quadrature angle~\cite{k2dimer,nicolas}. It happens because the $\chi^{(3)}$ interaction itself also changes the quadrature angles at which the optimal correlations may be measured~\cite{vova}, acting in some sense as an intensity dependent detuning~\cite{granja}. In principle it may be possible to dynamically control the detuning, coupling and intensities so that the largest violations of the inequalities were found for $\theta=0$, although we will not investigate this further because it is relatively simple to control the phase of local oscillators to choose the appropriate angles.     
We have shown the results for the quadrature angles at which the violation of the inequalities is maximised for different values of the nonlinear interactions. The existence of the maximal violations at finite frequencies could be a real operational advantage, due to the excess noise which is found experimentally at low frequencies. 

An alternative to indicate the existence of bipartite entanglement is a demonstration of the EPR paradox, which, as proven by Reid~\cite{mdr2}, is sufficient to demonstrate inseparability of the system density matrix. The EPR paradox itself was introduced in an attempt to show that quantum mechanics was not a complete and locally realistic theory~\cite{EPR}. Schr\"odinger replied by introducing the concept of entangled states which were not compatible with classical ideas such as local realism~\cite{gato}. In 1989 Reid proposed a physical test of the EPR paradox using optical quadrature amplitudes~\cite{mdr1}, which are mathematically identical to the position and momentum originally considered by EPR. Reid later expanded on this work, demonstrating that the satisfaction of her two-mode 1989 criterion, and hence a demonstration of EPR correlations, always implies bipartite quantum entanglement~\cite{mdr2}. It was also shown by Tan~\cite{Sze} that the existence of two orthogonal quadratures, the product of whose variances violates the limits set by the Heisenberg uncertainty principle, provides evidence of entanglement. Tan demonstrated this in the context of teleportation, with the outputs from a nondegenerate OPA mixed on a beamsplitter.

\begin{figure}[tbhp]
\includegraphics[width=0.8\columnwidth]{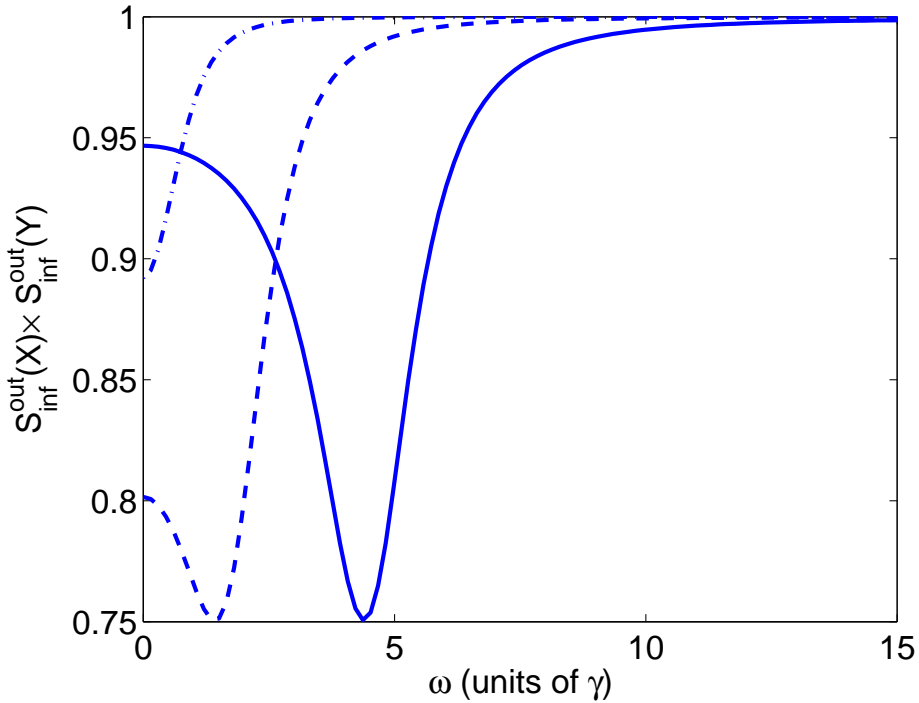}
\caption{Violation of the EPR inequality for $\epsilon=10^{3},\gamma=1,J=\Delta=10$. The solid line is for $\chi=10^{-5}$, at a quadrature angle of $\theta=80^{\rm o}$, the dashed line is for $\chi=10^{-6}$, at an angle of $\theta=122^{\rm o}$ and the dash-dotted line is for $\chi=10^{-7}$, at an angle of $\theta=14^{\rm o}$.}
\label{fig:EPR}
\end{figure}

To examine the utility of the system for the production of states
which exhibit the EPR paradox, we will follow the approach of Reid~\cite{mdr1}, as outlined in Dechoum \etal~\cite{ndturco}.
We assume that a measurement of the $\hat{X}^{\theta}_{1}$ quadrature, for
example, will allow us to infer, with some error, the value of the
$\hat{X}_{2}^{\theta}$ quadrature, and similarly for the $\hat{Y}_{j}^{\theta}$ quadratures.
This allows us to make linear estimates of the quadrature variances,
which are then minimised to give the inferred variances, 
\begin{eqnarray}
V^{inf}(\hat{X}_{1}^{\theta}) &=& V(\hat{X}_{1}^{\theta})-\frac{\left[V(\hat{X}_{1}^{\theta},\hat{X}_{2}^{\theta})\right]^{2}}{V(\hat{X}_{2}^{\theta})}, \\
V^{inf}(\hat{Y}_{1}^{\theta}) &=& V(Y_{1}^{\theta})-\frac{\left[V(\hat{Y}_{1}^{\theta},\hat{Y}_{2}^{\theta})\right]^{2}}{V(\hat{Y}_{2}^{\theta})},
\label{eq:EPROPA}
\end{eqnarray}
where $V(A,B)=\langle AB\rangle-\langle A\rangle\langle B\rangle$.
The inferred variances for the $j=2$ quadratures are simply found
by swapping the indices $1$ and $2$. As the $\hat{X}_{j}^{\theta}$ and $\hat{Y}_{j}^{\theta}$
operators do not commute, the products of the actual variances obey a Heisenberg
uncertainty relation, with $V(\hat{X}_{j}^{\theta})V(\hat{Y}_{j}^{\theta})\geq 1$.
Hence we find a demonstration of the EPR paradox whenever 
\begin{equation}
V^{inf}(\hat{X}_{j}^{\theta})V^{inf}(\hat{Y}_{j}^{\theta})<1.
\label{eq:demonstration}
\end{equation}
In Fig.~\ref{fig:EPR} we show results for the product of the inferred spectral output variances, indicating a clear demonstration of the EPR paradox and hence entanglement between the modes. The violations are found to be maximum for the same quadrature angles as found for the Duan inequalities, with a qualitative similarity between the results. The small violations at higher frequencies, shown in Fig.~\ref{fig:Duan} for the Duan inequalities, were not so apparent for the products of the inferred variances. This can be understood because the Duan criterion provides necessary and sufficient conditions for entanglement with a Gaussian system, while the EPR correlation provides a sufficiency criterion. Operationally, it would not be a problem as the major violations at lower frequencies are potentially more interesting, due to the higher intensities of the output fields closer to the cavity resonance frequency. 

\begin{figure}[tbhp]
\includegraphics[width=0.8\columnwidth]{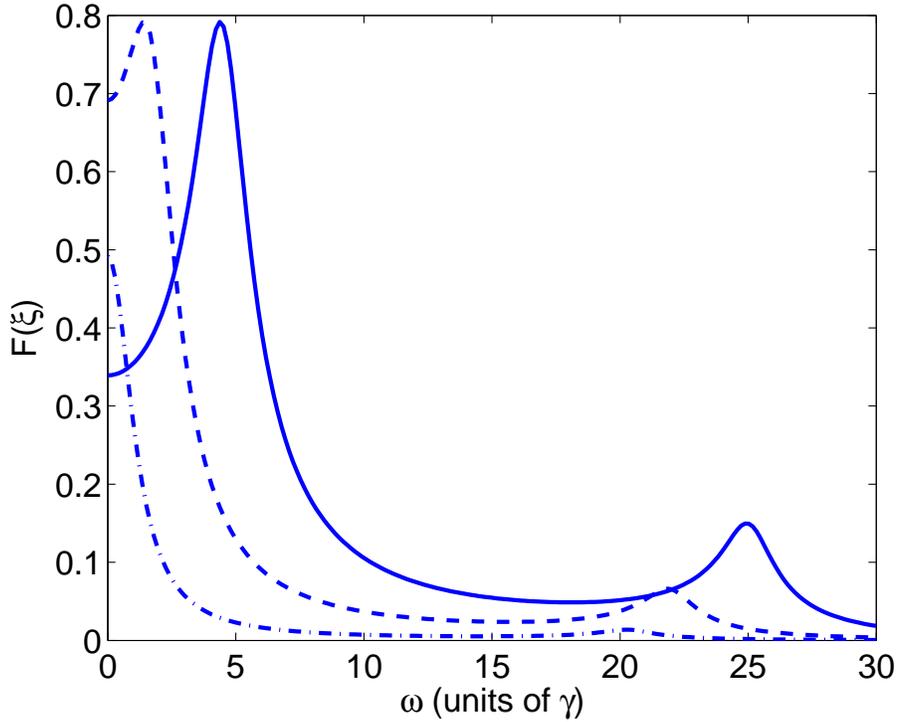}
\caption{The logarithmic negativity as a function of frequency for $\epsilon=10^{3},\gamma=1,J=\Delta=10$. The solid line is for $\chi=10^{-5}$, the dashed line is for $\chi=10^{-6}$ and the dash-dotted line is for $\chi=10^{-7}$.}
\label{fig:lneg}
\end{figure}

The third measure which we will apply is the logarithmic negativity, proposed by Vidal and Werner as a computable measure of entanglement, as opposed to others which can be difficult to calculate~\cite{Goofrey}. This measure has been used by Mancini to quantify the effects of feedback on continuous variable entanglement in a two-mode system~\cite{stefano}. We note here that this measure is defined for Gaussian states, to which we are also limited here due to the linearisation process we have used. We first define the system covariance matrix as
\begin{equation}
{\cal C}=\left[\begin{array}{cc}
{\cal C}_{1} & {\cal C}_{12} \\
{\cal C}_{21} & {\cal C}_{2}
\end{array}\right],
\label{eq:covmat}
\end{equation}
where
\begin{equation}
{\cal C}_{j}=\left[\begin{array}{cc}
V(\hat{X}_{j}^{\theta}) & V(\hat{X}_{j}^{\theta},\hat{Y}_{j}^{\theta}) \\
V(\hat{Y}_{j}^{\theta},\hat{X}_{j}^{\theta}) & V(\hat{Y}_{j}^{\theta})
\end{array}\right],
\label{eq:jcovmat}
\end{equation}
and
\begin{equation}
{\cal C}_{ij}=\left[\begin{array}{cc}
V(\hat{X}_{i}^{\theta},\hat{X}_{j}^{\theta}) & V(\hat{X}_{i}^{\theta},\hat{Y}_{j}^{\theta}) \\
V(\hat{Y}_{i}^{\theta},\hat{X}_{j}^{\theta}) & V(\hat{Y}_{i}^{\theta},\hat{Y}_{j}^{\theta})
\end{array}\right].
\label{eq:ijcovmat}
\end{equation}
Defining
\begin{equation}
\xi = \sqrt{(\det{\cal C}_{1}-\det{\cal C}_{12})-\sqrt{(\det{\cal C}_{2}-\det{\cal C}_{12})^{2}-\det{\cal C}}}
\end{equation}
the logarithmic negativity is then defined as
\begin{displaymath}
{\cal F}(\xi) = \left\{\begin{array}{ll}
-\log_{2}\xi & \textrm{if $\xi<1$}\\
0 & \textrm{otherwise.}
\end{array}\right.
\end{displaymath}
Any non-zero value of ${\cal F}(\xi)$ is then an indication that the two modes are entangled. An interesting feature of this measure is that it has no dependence on quadrature angle, at least with the system we are investigating here. The values shown in Fig.~\ref{fig:lneg} are valid for all angles, even though the other two measures will not indicate entanglement for arbitrary angle. This shows that the logarithmic negativity is useful for demonstrating that continuous variable entanglement exists in a given system, but does not tell us at which quadrature angles the system may exhibit the necessary properties for uses such as teleportation. We note that the frequencies at which the logarithimic negativity is non-zero are the same as those at which the Duan inequalities are violated. 

\section{Conclusions}

We have analysed the intracavity Kerr nonlinear coupler in terms of its entanglement properties and shown that it can produce macroscopically intense outputs which possess continuous variable entanglement. This entanglement has been demonstrated by the calculation of three different measures, the first two of which give maximum violations of the appropriate inequalities at particular quadrature angles, and the third of which indicates the presence of entanglement in a manner which is not phase dependent. As this device could be constructed using integrated optics and only needs to be optimised for one frequency of operation, it could provide a rugged and stable source of bright entangled beams which could be of some utility for such technological uses as, for example, quantum communication channels and continuous variable teleportation.

\begin{acknowledgments}

This research was supported by the Australian Research Council.

\end{acknowledgments}

\end{document}